\documentclass[twocolumn,showpacs,amsmath,amssymb,prl]{revtex4}

\usepackage{graphicx}
\usepackage{dcolumn}
\usepackage{bm}

\begin{document}

\title{Measurement of fundamental thermal noise limit in a cryogenic sapphire frequency standard using bimodal maser oscillations}
\author{Karim Benmessai$^1$}
\author{Daniel Lloyd Creedon$^2$}
\author{Michael Edmund Tobar$^{2}$}
\email{mike@physics.uwa.edu.au}
\author{Pierre-Yves Bourgeois$^1$}
\author{Yann Kersal\'e$^1$}
\email{yann.kersale@femto-st.fr}
\author{Vincent Giordano$^1$}
\affiliation{\\ $^1$Institut FEMTO-ST, UMR 6174 CNRS, Universit\'e de Franche, Comt\'e, 25044 Besan\c{c}on, France\\
$^2$University of Western Australia, School of Physics M013, 35 Stirling Hwy., Crawley 6009 WA, Australia}

\date{\today}

\begin{abstract}
We report observations of the Schawlow-Townes noise limit in a cryogenic sapphire secondary frequency standard. The effect causes a fundamental limit to the frequency stability, and was measured through the novel excitation of a bimodal maser oscillation of a Whispering Gallery doublet at $12.04 GHz$. The beat frequency of $10 kHz$ between the oscillations enabled a sensitive probe for this measurement of fractional frequency instability of $10^{-14}\tau^{-1/2}$ with only $0.5$ $pW$ of output power.
\end{abstract}

\maketitle

Cryogenic sapphire oscillators are based on single cylindrical crystals of sapphire of approximately $50 mm$ in diameter, cooled with liquid helium and operated by exciting Whispering Gallery modes with $Q$-factors of order $10^9$ at microwave frequencies \cite{chang97}. State-of-the-art stabilities of parts in $10^{16}$ ($\Delta f/f<10^{-15}$) have been achieved for integration times between 2 to 1000 seconds \cite{hartnett, Chang}. This has lead to a range of unique applications including precision tests of Lorentz Invariance, such as Michelson-Morley \cite{MM, Stanwix, StanwixPRD}, Kennedy Thorndike \cite{KT, WolfGRG, wolfprl} and Standard Model Extension \cite{Stanwix, StanwixPRD, Kosto1_1, Kosto1_2, Kosto1_3, KM} experiments. Other applications include quantum-limited operation of laser cooled atomic frequency standards. The high stability of the sapphire oscillator allows the elimination of technical noise due to the pulsed sampling required to generate Ramsey fringes \cite{giorgio}. To date, cryogenic oscillators are the only electromagnetic oscillators with sufficient stability to achieve this. Thus, many laboratories worldwide are incorporating them into their clock ensembles to generate microwave and optical signals for such purposes \cite{hartnett, Watabe, Daussy, McFerran, Bourgeois, Fisk, Femto}. As well as the practical applications with regards to timing, navigation and space applications \cite{Salomon}, the fountain clocks have now achieved sufficient sensitivity to test the stability of fundamental constants \cite{Marion, Bize, Tobar1}.

The fundamental limits of frequency stability in such high precision oscillators are due to competing effects from fluctuating radiation pressure of the electromagnetic field or small random background fluctuations due to thermal \cite{Nyquist} or quantum Nyquist noise processes (Schawlow-Townes noise limit)  \cite{Oliver, Brag}. These limits represent a barrier to frequency instability that cannot be surpassed using classical techniques. It is important to characterise these noise processes through precise measurement. For a cryogenic sapphire oscillator operating at the usual levels of carrier power ($10 mW$), radiation pressure effects have already been characterised \cite{chang97}. In contrast, the frequency instability due to the Schawlow-Townes noise limit for a cryogenic sapphire oscillator is of order $10^{-20}$ at $1$ second averaging time at this carrier power. In this regime the performance of precise frequency generation is limited by technical noise sources such as flicker noise in the amplification process of the oscillator sustaining stage (solid-state amplifier). Typically this is at a level of parts in $10^{16}$. In this work we report on the first observation of the fundamental frequency stability limit due to Nyquist thermal fluctuations in a Cryogenic Sapphire Oscillator using a three-level zero-field maser transition as the oscillator sustaining stage.

The fundamental spectral density of Nyquist noise of an electromagnetic mode sustained by an ideal amplifier is given by Eqn. (\ref{Nnoise}) \cite{Oliver}. 
\begin{equation}
\psi_a(\nu)=\frac{h \nu}{e^{\frac{h \nu}{k_B T}}-1}+\frac{h \nu}{2} \ \ [W/Hz]   \label{Nnoise}
\end{equation}
Here, $h$ is Planks constant, $\nu$ the frequency of the mode, $k_B$ Botzmanns constant and $T$ the temperature of the surrounding environment. The thermal regime is distinguished from the quantum regime for mode frequencies, $\nu$ $<$ $k_B/h$ , which is $100 GHz$ at liquid helium temperature. In contrast to laser technology our experiment is in the Nyquist thermal regime and the Schawlow-Townes limit on frequency instability for this case is given in Eqn. (\ref{ST}) \cite{Schawlow}.
\begin{equation}
\sigma_y(\tau)=\frac{1}{Q_{mode}}\sqrt{\frac{k_B T}{2P_{mode}\tau}}    \label{ST}
\end{equation}
Here, $Q_{mode}$ is the mode $Q$-factor, $P_{mode}$ is the power in Watts, and $\tau$ is the measurement integration time in seconds.

Recently a new type of cryogenic sapphire oscillator was discovered based on masing from residual impurity ions with zero applied magnetic field \cite{APL}. Maser action results from the coincidence in frequency of the Electron Spin Resonance of Fe$^{3+}$ ions with a very high $Q (>10^9)$ Whispering Gallery Mode at $12.04 GHz$. The output power obtained with this device is of order $2.5$ $nW$, which is ten thousand times higher than a hydrogen maser. This leads to the fundamental Schawlow-Townes frequency stability limit due to Nyquist thermal fluctuations acting on the Whispering Gallery Mode resonator of order $10^{-16} \tau^{-1/2}$. The first characterization against the state-of-the-art classical cryogenic sapphire oscillator (solid-state sustaining stage) resulted in the measurement of an upper limit to the frequency instability of two orders of magnitude higher of order $10^{-14}$ at $1$ second, which was limited by the microwave frequency synthesis chain used for the frequency comparison \cite{pybuwa}. 

\begin{figure}
\begin{center}
\includegraphics[width=3in]{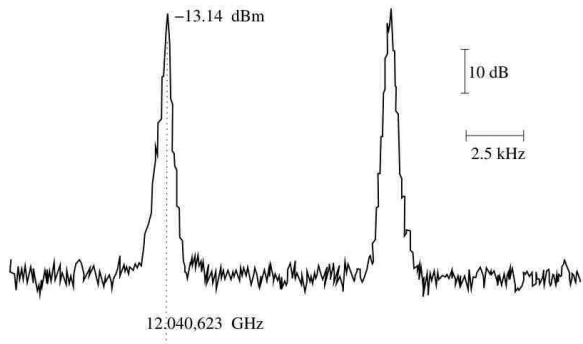}
\caption{Typical spectrum of the bimodal oscillation at 12.04 GHz using a readout with 80 dB gain.} \label{bimode}
\end{center}
\end{figure}

Whispering Gallery Modes in such high-$Q$ resonators in reality exist as doublets due to perturbations such as backscatter in the crystal due to Rayleigh scattering and probe perturbations \cite{wg1,wg2,wg3,splitting}. The back scatter causes the two opposite traveling waves propagating around the circumference of the sapphire cylinder to exhibit strong coupling and split into two spatially orthogonal circumferential standing waves \cite{wg2}, which in this case are separated by 10 kHz in frequency. Because the splitting is small ($\approx10 kHz$) it is only observable in high-Q systems. In this work we report simultaneous maser action of the two modes as shown in Fig.\ref{bimode} (bimodal operation). In contrast, the bimodal regime only occurs at low power levels between $10^{-15}$ to $10^{-12}$ $Watt$ output power (at higher powers only one mode dominates). For these power levels the Schawlow-Townes frequency instability limit is larger and of order $10^{-14}\tau^{-1/2}$. In this work, we show that the mixing of the two bimodal signals has enabled us for the first time to measure this limit for a cryogenic sapphire oscillator, and verified the low noise properties of the Maser.

\begin{figure}
\begin{center}
\includegraphics[width=3in]{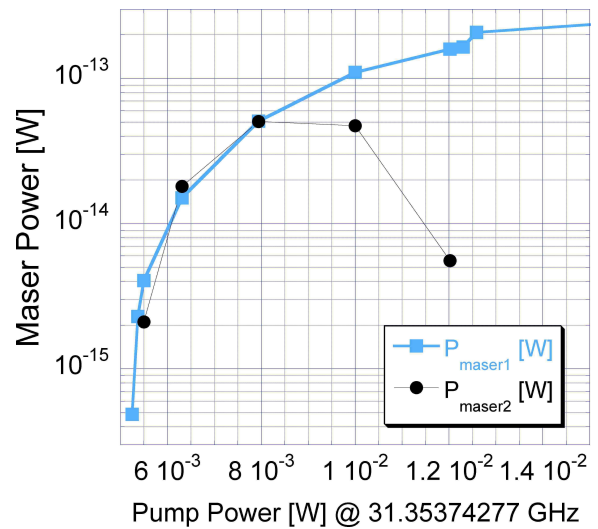}
\caption{Typical output power of the bimodal maser oscillation as a function of pump power for the 31.35374277 GHz pump mode. The actual output power depends on the mode chosen to pump the mode and whether or not a magnetic field is applied} \label{fig:1}
\end{center}
\end{figure}

The resonator consists of a HEMEX sapphire single crystal disk 50mm diameter and 30mm high mounted in a gold plated OFHC copper cavity. The operating mode is the $WGH_{17,0,0}$ at $12.04 GHz$ cooled to liquid helium temperature with a pulsed tube cryocooler. The doublet pair, separated by $10 kHz$, is characterized by a loaded $Q$-factor of the order of $7\times10^8$ at $4.2K$. To create the population inversion we inject in the resonator a signal between $31.305$ to $31.411$ $GHz$, via a microwave synthesizer, corresponding to the center frequency of another high-$Q$ Whispering Gallery mode. There are 34 modes in this frequency range that can be excited to create a population inversion in the gain medium, which in turn causes the 12.04 GHz mode to oscillate. By applying this pump signal with a power above the threshold ($\approx10 mW$) and within this 100 MHz range, the two maser signals may coexist at the resonator output probe. A typical plot of the bimodal maser signal power versus pump power is shown in Fig.\ref{fig:1}. Only in the low power regime can both modes co-exist (similar to two-mode operation of a ring laser) when coefficients due to cross saturation and self-saturation between the two coupled-modes result in positive net gain for both modes \cite{rlaser}. In this case population inversion is created mainly from separate parts of the maser medium. In contrast, in the high power regime only one mode can oscillate due to competition across the whole maser medium. On the transition from the bimodal to single-mode operation (Fig.\ref{fig:1}) there is a discrete jump in masing power of the oscillating mode, due to the discrete change in medium gain.

\begin{figure}
\begin{center}
\includegraphics[width=3in]{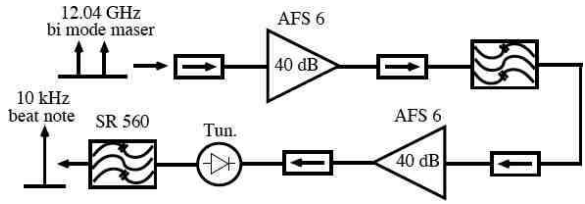}
\caption{Beat frequency readout for the bimodal maser.}
\label{fig2}
\end{center}
\end{figure}

The two signals, extracted from the same probe, are amplified by an amplifier chain (two amplifiers with $40 dB$ gain each) and then sent to a quadratic detector (tunnel diode) to be mixed (Fig.\ref{fig2}). A microwave band pass filter is placed between the two microwave amplifiers to reduce the thermal noise produced by the first amplifier. The $10 kHz$ beat note is band pass filtered and amplified by a low noise amplifier and then sent to a high-resolution counter, referenced to a hydrogen maser, to extract the frequency stability.

\begin{figure}
\begin{center}
\includegraphics[width=3in]{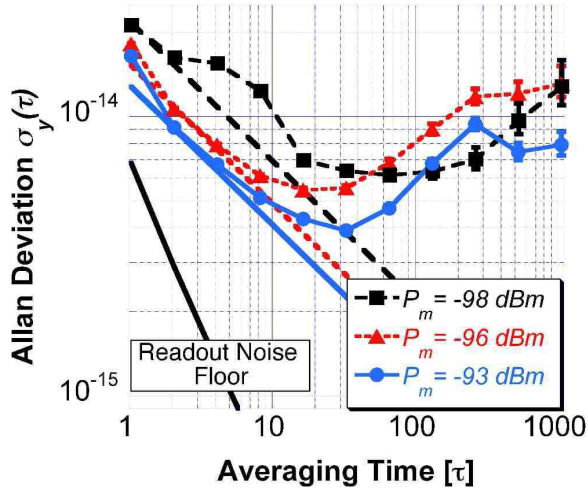}
\caption{Allan deviation versus integration time for various power levels in units of $dBm$ ($dB$ with respect to a $mW$). The readout noise floor is calculated from the measured phase noise, using the transformations of white and flicker phase noise presented in \cite{sam}. The white noise level is calculated from the tangents with $\tau^{-1/2}$ dependence. At $1$ second the stability is artificially raised by a periodic signal due to the cryocooler at $0.45 Hz$ (see Fig.\ref{fig:stab}). Periodic effects are also visible in the $P_{mode} = -98 dBm$ curve at $0.67 Hz$ (bump at $2-3$ seconds) and $P_{mode}= -93 dBm$ curve at $1.3 mHz$ (bump at $300$ seconds).}
\label{fig:stab}
\end{center}
\end{figure}

\begin{figure}
\begin{center}
\includegraphics[width=3in]{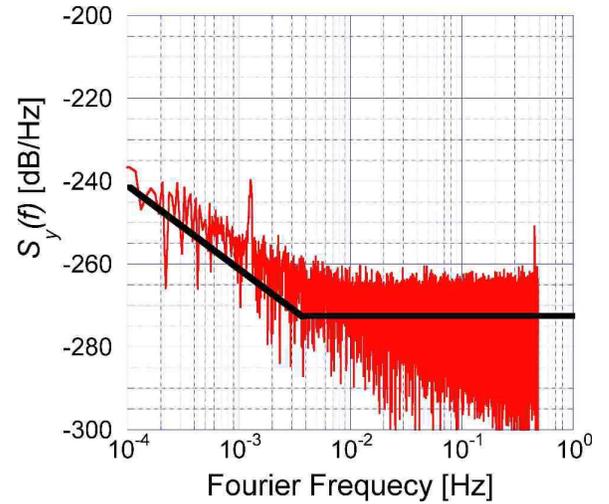}
\caption{Frequency noise spectral density of $P_{mode} = 0.5 pW$, showing the white noise floor. The spikes in the spectrum at $0.45 Hz$ and $1.3 mHz$ manifest in the Allan deviation as bumps at $1$ and $300$ seconds respectively in Fig.\ref{fig:stab}.}
\label{fig4}
\end{center}
\end{figure}

The frequency instability in terms of Allan deviation is plotted in Fig.\ref{fig:stab} for various maser output powers. The level of white noise is calculated from the coefficient of the $\tau^{-1/2}$ fit in the $1$ to $10$ second interval. Periodic effects cause clearly visible bumps in the Allan deviation curve, which can be ignored when calculating the white noise. From the spectral density of fractional frequency fluctuations plotted in Fig.\ref{fig4}, the periodic effects are clearly visible, and for Fourier frequencies $> 10^{-2} Hz$ the white noise floor is clearly visible.

\begin{figure}
\begin{center}
\includegraphics[width=3in]{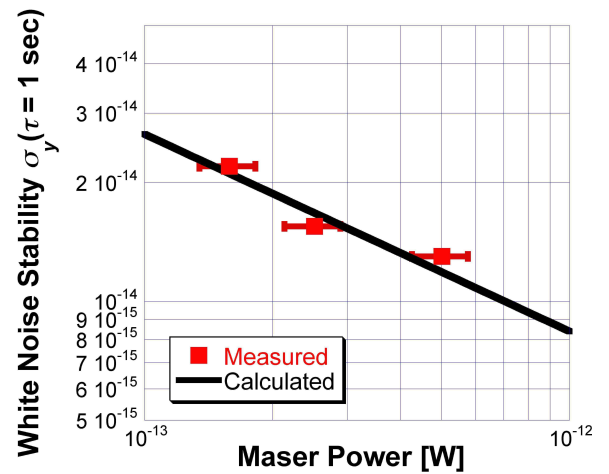}
\caption{Value of white noise component at $1$ second as deduced from Fig.\ref{fig:stab}. The points with error bars are the measured values, while the solid curve is the value calculated from Eqn.(\ref{ST}).}
\label{fig5}
\end{center}
\end{figure}

The output power of the maser is measured on the spectrum analyser with a $40 dB$ gain amplifier. The error bars are calculated by taking ten measurements, then calculating the mean and standard deviation. The solid curve is the predicted stability assuming ideal amplification of the Maser using Eqn. (\ref{ST}). The calculation is done for a temperature of $5 K$, $Q$-factor of $7 \times 10^8$ and averaging time of $1$ second. From the comparison it is clear that we have measured the thermal noise limit given by the Schawlow-Townes limit on frequency stability.

In conclusion we report observations of thermally induced Nyquist noise due to temperature fluctuations in a high precision electromagnetic oscillator as calculated by the Schawlow-Townes formula. The measurements showed the correct dependence on power and showed that the maser action produced a sustaining stage that was close to ideal. This work suggest that a cryogenic sapphire oscillator operating with a maser sustaining stage in single mode operation ($2.5 nW$ output power) has the potential to improve the short term stability of cryogenic sapphire oscillators to a value of $10^{-16}\times \tau^{-1/2}$.

\begin{acknowledgments}
This work was supported by the Centre National d'\'Etudes Spatiales (CNES), the Agence Nationale pour la Recherche (ANR), the Australian Research Council and travelling support from FAST program from Egide and the International Science Linkages program from DEST.
\end{acknowledgments}


\begin{thebibliography}{99}
\bibitem{Femto} Although in the future microwave generation via an optical frequency standard and a femto-second comb may rival these results.
\end{thebibliography}
\end{document}